\documentclass[conference]{IEEEtran}
\ifCLASSINFOpdf
\else
\fi

\usepackage{color}
\usepackage{url}
\usepackage{graphicx}
\usepackage[]{algorithm2e}

\begin{document}
\title{AUGURY: A time-series based application for the analysis and forecasting
  of system and network performance metrics}

\author{\IEEEauthorblockN{Nicolas Gutierrez\IEEEauthorrefmark{1}\IEEEauthorrefmark{2}
  and Manuela Wiesinger-Widi\IEEEauthorrefmark{3}}
  \IEEEauthorblockA{\IEEEauthorrefmark{1}Institute for Particle Physics Phenomenology,
    Department of Physics, Durham University - DH1 3LE, United Kingdom\\
    \IEEEauthorrefmark{2}Department of Physics and Astronomy, University College London - London,
    WC1E 6BT, United Kingdom\\
    Email: nicolas.gilberto.gutierrez.ortiz@cern.ch
  }
  \IEEEauthorblockA{\IEEEauthorrefmark{3}
    RISC Software GmbH, Unit Advanced Computing Technologies\\
    IT-Center, Softwarepark 35, 4232 Hagenberg Austria\\
    Email: manuela.wiesinger@risc-software.at}
}

\maketitle

\begin{abstract}
  This paper presents AUGURY, an application for the analysis of monitoring data from computers,
  servers or cloud infrastructures.
  The analysis is based on the extraction of patterns and trends from historical data,
  using elements of time-series analysis.
  The purpose of AUGURY is to aid a server administrator 
  by forecasting the behaviour and resource usage of specific applications
  and in presenting a status report in a concise manner.
  AUGURY provides tools for identifying network traffic congestion
  and peak usage times, and for making memory usage projections.
  The application data processing specialises in two tasks: the parametrisation
  of the memory usage of individual applications
  and the extraction of the seasonal component from network traffic data.
  AUGURY uses a different underlying assumption for each of these two tasks.
  With respect to the memory usage, a limited number of single-valued parameters
  are assumed to be sufficient to parameterize
  any application being hosted on the server.
  Regarding the network traffic data, long-term patterns, such as hourly or daily exist
  and are being induced by work-time schedules and automatised administrative jobs.
  In this paper, the implementation of each of the two tasks is presented,
  tested using locally-generated data,
  and applied to data from weather forecasting applications hosted on a web server.
  This data is used to demonstrate the insight that AUGURY
  can add to the monitoring of server and cloud infrastructures.
\end{abstract}


% \fi
\IEEEpeerreviewmaketitle

\section{Introduction}

The PIPES-VS-DAMS framework~\cite{PvD} for the monitoring of cloud infrastructures
supervises various system and network performance metrics.
Vast amounts of this data are available for studying.
Analysing it and extracting regular patterns could help to identify hazardous trends.
These patterns can be used to plan and optimise the usage of resources,
and to highlight extraordinary events requiring detailed attention.
In this paper, AUGURY, an application for the extraction of such trends and patterns
in data from computers, servers or cloud infrastructures is presented.

AUGURY is able to handle various input data formats.
Internally, a unified data structure exists, which is transformed according to the use case, and
most importantly, for the application of time-series analysis techniques.
AUGURY contains queries for accessing subsets of the internal data structure
and for modifying it.
It also handles overlaps and gaps between input data sets.
The output is an aggregated data format, which is illustrated in this paper
using snapshots, representing the status of this output at some particular point in time.
%representing at some particular point of time the status 
%of this output, which evolves in time.

AUGURY uses time series analysis in two ways: to extract the parameters
for a model intended to represent the memory usage of a single application, and
to extract the seasonal component from the network traffic data.
%By building models for the memory usage of applications, it enables
%memory-usage projections requiring
%only a knowledge of the respective number of times
%these applications are executed within a given time window and summing them up.
Building models of the memory usage of single applications
enables projections of overall memory usage.
Forecasting the memory usage for a specific time window subsequently
only requires summing up the models for all application executions within that time frame.
By obtaining a seasonal component, a forecast
for that number of executions is also acquired.
Combining both, a network traffic congestion and peak usage times can be identified, studied
and expressed in terms of memory-usage projections.

A brief list of related work,
in the context of monitoring
and/or forecasting of system and network performance metrics, is presented in the following.
Refs.~\cite{NFS,Related1,Related2,Related3,Related4,Related5}
also approach this topic using time-series analysis,
but concentrate on the accuracy of the forecast.
In contrast, in this work, the focus is on the intuitiveness and usability
of the time-series analysis outcome, from the perspective of a system administrator.
A variety of network traffic monitoring tools
exist\footnote{\url{http://www.cs.wustl.edu/~jain/cse567-06/ftp/net_traffic_monitors3/}},
yet AUGURY is novel in that it focuses on the analysis and visualisation of seasonal patterns.

This paper is organised as follows.
First, in Section~\ref{sec:Data} the features of the data samples are discussed.
Then, in Section~\ref{sec:Theory}, a well-known seasonal-pattern extraction method is introduced
in parallel with the elements from time series analysis upon which it is based on.
In Section~\ref{sec:Modelling} the memory-usage model and parameter-extraction
algorithm are described.
Then, the specifications of AUGURY and the packages
it uses are discussed in Section~\ref{sec:implementation}.
The tests performed using locally-generated data are presented in Section~\ref{sec:LocalData}
and the results based on real-world data are shown in Section~\ref{sec:RealData}.
An auxiliary functionality provided by AUGURY for the modelling of residuals
is discussed in Section~\ref{sec:Residual}.
Finally, in Section~\ref{sec:Summary}, the conclusions are presented.

% Data
\section{Data samples}
\label{sec:Data}
Three data samples are used in this paper:
\begin{itemize}
\item memory-usage from a single application;
\item network-traffic from server requests with a fixed schedule;
\item monitoring of a web server.
\end{itemize}
The first two are used for testing and validation,
while the latter is used for establishing a use case for AUGURY.

Memory-usage data is collected from a virtual environment hosting a GNU/Linux system.
This emulates the conditions of an isolated application.
Regular executions of a task with a rectangle-like memory
usage pattern are invoked using crontab.
A lapse between executions of two minutes is used.
The virtual machine performance metrics are monitored
using \textsc{glances}\footnote{\url{https://nicolargo.github.io/glances/}}, an open source
software.
The status is exported every two seconds to a Comma-Separated Values (CSV) file.
This file is shared with the host machine, where the data is analysed.
Several performance metrics are exported into the CSV file,
including memory and CPU usage.

Network traffic data is collected from an apache server installed on a Raspberry Pi computer.
A second Raspberry Pi sends requests to this server via a local network.
A script submitting these requests is scheduled to run every minute.
%The number of request depends only on the minutes and, hence,
Thus, the network traffic has an inherent seasonal pattern.

Web-server monitoring data is collected from a server hosting weather forecasting applications.
A limited number of applications dominate this sample.
The 5 most frequently used applications make for approximately 70$\%$ of the total traffic.
Each of these applications corresponds primarily to a single IP-address submitting the requests.

\section{Seasonal adjustment}
\label{sec:Theory}

In this section, the procedure used to extract the seasonal
component from a dataset is introduced.
AUGURY's main use-case is the study of this component.

A time series~\cite{TimeSeries1,TimeSeries2,TimeSeries3}
is a sequence of time-dependant data points,
where the lag between them is constant in time.
Time series are studied mainly
to understand the time-dependant behaviour of some quantity,
for instance, fluctuations in the price of some stock or seasonal patterns in the demand
for a certain product.
An application of time series analysis is the forecasting of the future value
of such a quantity based on previous observations. That is
\begin{equation}
y_{t+1} = F(y_t,y_{t-1},....,y_{t-(N-1)}),
\end{equation}
where $t$ denotes the time of the last (most recent) out of $N$ observed values
of the variable $y$.
This function is by no means known, and could depend on derivative variables,
such as a Moving Average (MA) or standard deviation ($\sigma$),
and on stochastic terms.

The MA can be defined in multiple ways, according to the weights assigned to each
of the lagged values.
In general, any MA is given by:
\begin{equation}
  \mathrm{MA}(N,w)_t = \frac{w_t*y_t + \dots + w_{t-(N-1)}*y_{t-(N-1)}}{\sum{w_i}},
  \label{eq:MA}
\end{equation}
where $N$ corresponds to the number of lags used and $w_i$
is the weight given to each observation.
The standard case corresponds to $w_i=1$ for any $i$.
Other versions are also frequently used, for instance, the Exponentially Weighted MA (EWMA)
where the largest weight is given to the most recent observation.
Here, the weight decays exponentially as the observations move further back in time.

The calculation of the MA requires at least $N$ observations and, therefore, no value of the MA
exists for the first $N-1$ points.
In the extraction of the seasonal pattern a symmetric MA is used,
i.e. the same number of past and future observations is used, and, hence,
this procedure yields missing values at both the beginning and the end of the dataset.

A seasonal pattern is one which repeats periodically, i.e. there is a period $p$ for which:
\begin{equation}
S_{t+p} = S_t,
\end{equation}
where $S_t$ is the seasonal component of a time series.
In this paper, three types of seasonality are considered,
labelled as hourly, daily or weekly depending on whether $p$
is an hour, day or week, respectively.

In this work, the extraction of seasonal patterns is based on sample decomposition.
That is 
\begin{equation}
Y_t = S_t + T_t + E_t
\end{equation}
where the series $Y_t$ is given by the combination of the seasonal ($S_t$),
trend ($T_t$) and the stochastic error ($E_t$) terms.
Established methods for such a decomposition exist and are documented
elsewhere~\cite{STL}.
%~\cite{SA_wiki,X13}.
These are typically referred to as \textit{seasonal adjustment}~\cite{SeasonalAdjustment}
and are based on the application of a chain of filters.
First, a MA is used to estimate the trend component, since
changes on a smooth MA can be associated to a trend.
This detrended, \textit{filtered}, data is then used to extract the seasonal and error terms.

\section{Memory usage parametrisation}
\label{sec:Modelling}

In this section, the model used to characterise the memory usage of a single application,
and the strategy devised to estimate its parameters, is presented. The purpose of this model
is to be used in memory-usage projections.

\begin{figure}[!hbt]
  \begin{center}
    \includegraphics[width=\columnwidth]{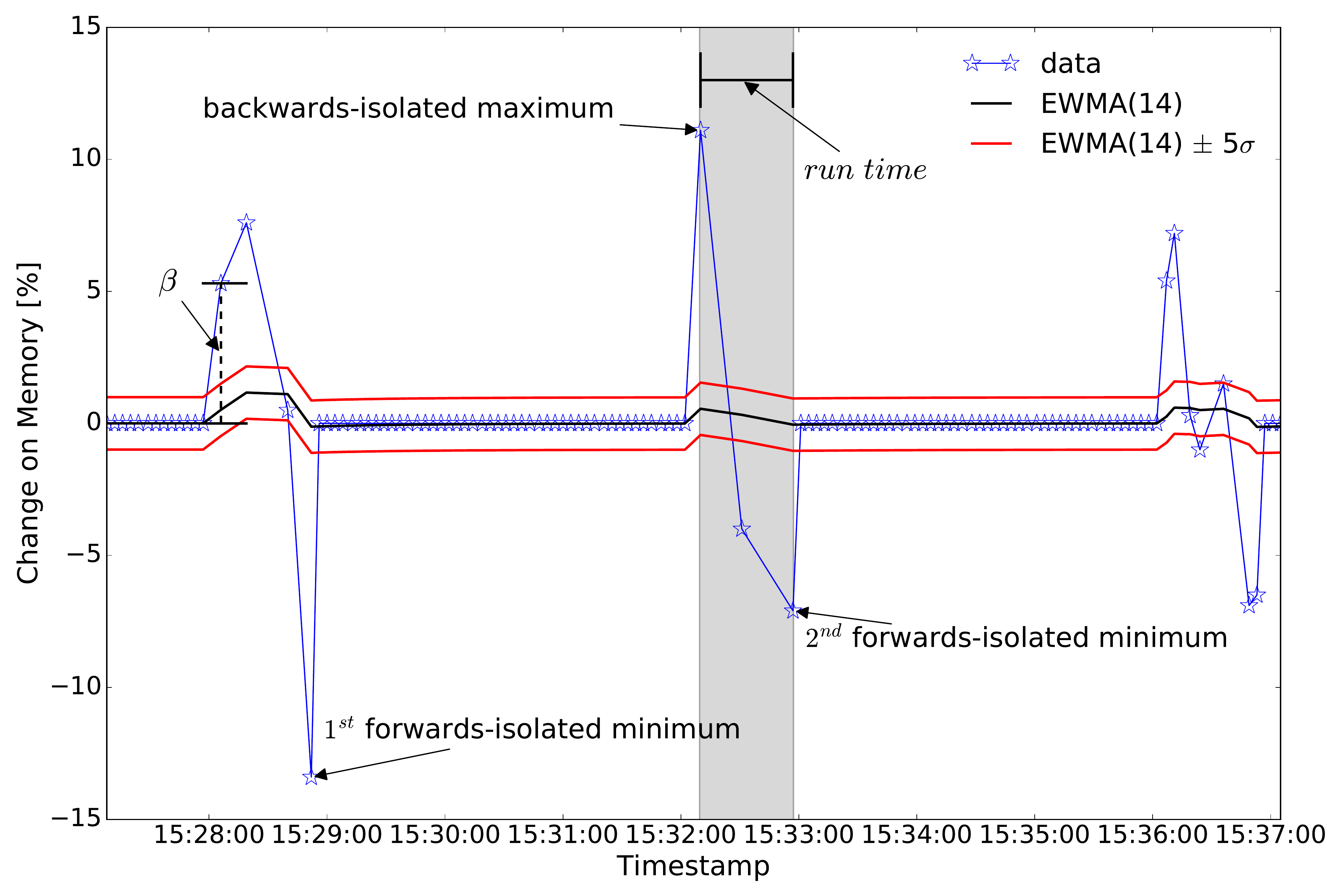}
    \caption{Change on the percentage of memory used ($y'_t$)
      for a time window of approximately 10 minutes. 
      The backwards-isolated maximum and the $2^{nd}$ forwards-isolated minimum constitute
      a signal-like pattern.
      The $1^{st}$ forwards-isolated minimum signals the end of the previous impulse.
      Also here, the model parameters $\beta$,  corresponding to the magnitude of
      the first significant deviation,
      and $run\ time$ are illustrated.
    }
    \label{fig:SignalPattern}
  \end{center}
\end{figure}

A simple three-parameter model is used to characterise the memory usage of a given application.
These are the running time ($run\ time$), the maximum memory used ($max\ memory$) and $\beta$,
the latter of which is illustrated in Fig.~\ref{fig:SignalPattern}
and corresponds to the magnitude of the first significant deviation.

The extraction of the model parameters relies on the identification of signal-like patterns.
These structures are recognised by transforming the series as
\begin{equation}
y'_t = y_t - y_{t-1}.
\end{equation}
This transformation is similar to a derivative, in that sudden changes in $y_t$ spawn
a local maximum or minimum in $y'_t$,
while plateau regions, where $y_t \approx y_{t-1}$ and hence $y'_t \approx 0$, are relatively flat.
This is illustrated in Fig.~\ref{fig:SignalPattern}.
Here, $y'_t$ is shown for a chain of signal impulses with a rectangular-like shape.
The pattern-recognition strategy is presented in the following.

\subsection{Statistical significance of a deviation}

Significant deviations, minima or maxima, are identified using a MA (see Eq.~\ref{eq:MA}).
In this case, their significance is calculated with respect to the
standard deviation of the MA ($\sigma_{\mathrm{MA}}$).
A deviation must be outside the MA$\pm 5\sigma_{\mathrm{MA}}$ region to be identified as significant.
This is illustrated in Fig.~\ref{fig:SignalPattern}.
Here, a 14-lag MA (black line) and the $\pm 5\sigma_{\mathrm{MA}}$ band (red lines)
are shown on top of the $y'_{t}$ data.

The choice of the $\sigma$ scaling factor expresses the degree of belief
in the background (noise) model.
According to Chebyshev's inequality~\cite{Tchebichef} 96$\%$ of the
background distribution is between $\mu \pm 5 \sigma$, with $\mu$ being the mean,
regardless of the distribution.
By requiring a deviation to be further than $5\sigma$ from the MA to be labelled as significant,
the likelihood for the noise to fake a signal is, therefore, considerably minimised.

\subsection{Optimal moving average definition}

An EWMA is used to minimise the effect of the gap between signal
pulses in the dataset.
This foresees a real-life application, where the extent of this gap can fluctuate arbitrarily.
The resilience of the EWMA to the gap length can be easily illustrated by the following example.
Take a sudden change in $y'_t$, occurring at some given time $t$,
which is preceded by a gap of length $N-1$ where,
by definition, $y'_{t-1} \approx \dots \approx y'_{t-(N-1)} \approx 0$.
The MA in this case would be
\begin{equation}
  \mathrm{MA}(N,w)_t \approx \frac{w_t*y'_t}{\sum w_i}.
\end{equation}
For the standard case the MA yields $y'_t/N$,
in which case the MA drops as $N$ increases.
In contrast, the EWMA is closer to $y'_t$, as $w_t \approx \sum w_i$, and,
hence, has a negligible dependence on the gap length.

\subsection{Optimisation of the moving-average parameter $N$}

The MA has a free parameter, $N$.
This parameter has a direct impact on the labelling of significant deviations on $y'_t$.
On the one hand, a very large value of $N$ would most likely yield a smooth MA.
In this case, any deviation is significant and, therefore, the number of maxima-minima
pairs would be unmanageable, particularly for high levels of noise.
On the other hand, a small value of $N$ would yield a MA that fluctuates as often as $y'$ does.
In this case, $\sigma_{\mathrm{MA}}$ tends to be large and
signal-initiated deviations are likely to be mislabelled as noise. 
Considering all of the above, the MA parameter $N$ is chosen such that it simultaneously
minimises the number of significant deviations and $\sigma_{\mathrm{MA}}$.

An optimised value of the parameter $N$
is estimated using a discriminant $d_{i}$, where $i$ corresponds to each of
the values of $N$ being tested,
given by
\begin{equation}
  d_{i} = n_{5\sigma,i} + \sigma_{\mathrm{MA}(i,w)},
\end{equation}
where, $i$ runs in between 2 and the total number of observations available
and $n_{5\sigma,i}$ is the number of $5\sigma$ deviations of $y'$ from $\mathrm{MA}(i,w)$.
Also here, both $n_{5\sigma,i}$ and $\sigma_{\mathrm{MA}(i,w)}$
are normalised to their respective maximum values.
Therefore, $n_{5\sigma,i}$ and $\sigma_{\mathrm{MA}(i,w)}$ range between 0 and 1.
The value of $i$ that minimises $d_i$ is chosen as the optimal value of $N$
for the calculation of the MA.

\subsection{Extraction of the model parameters}

In order to extract the model parameters from the data,
the signal-like patterns must be found.
Such patterns are illustrated in Fig.~\ref{fig:SignalPattern}
by the backwards-isolated maximum and the $2^{nd}$ forwards-isolated minimum.
A maximum (minimum) is backwards-isolated (forwards-isolated)
if there are no other maxima (minima) within the last (next) three lags.
For each such pattern, the model parameters are estimated as:
\begin{itemize}
\item $\beta$: the magnitude of the maximum $y'_t$,
\item $max\ memory$: maximum value of $y_t$ within the pattern,
\item $run\ time$: length of the pattern.
\end{itemize}
The algorithm for finding the signal-like patterns proceeds as follows.
First the time window is divided in intervals delimited by pairs of
forwards-isolated minimums.
Then for every such interval a check of whether it contains a backwards-isolated maximum is made.
If it does, this maximum is paired with
the forwards-isolated minimum at the end of its interval.
The resulting pair indicates a signal-like pattern.
If it does not contain a backwards-isolated maximum, there is no signal-like
pattern in this interval.

\section{Implementation}
\label{sec:implementation}

AUGURY is implemented on \textsc{Python} and uses
%The unified data structure within AUGURY uses
the \textsc{Pandas}\footnote{\url{https://pypi.python.org/pypi/pandas}} package,
which provides a flexible data structure and dedicated functionality
for time series manipulation and statistical analysis.
The data processing is mostly based on list and dictionary comprehension
methods\footnote{\url{http://blog.endpoint.com/2014/04/dictionary-comprehensions-in-python.html}}
and, therefore, it does not show any sign of slowing down for very large datasets.

The data-reading interface is built on top of \textsc{Pandas}.
It has an interface for reading CSV files and is therefore capable to
read the memory-usage data from a single application.
The network-traffic data from scheduled server requests is available in the form
of \textsc{apache}\footnote{\url{https://www.apache.org/}} log files.
This is parsed into the CSV format.
The cloud-server monitoring data is in the JSON\footnote{\url{http://json.org/}} format,
which \textsc{Pandas} has an interface for.

The \textsc{statsmodels}\footnote{\url{http://statsmodels.sourceforge.net/}}
module for \textsc{Python}
is used for the seasonal adjustment.
Additional \textsc{Python} modules such as \textsc{numpy} and \textsc{datetime}
are used for numerical calculations and histogram-building, and time-stamp manipulation,
respectively.
The figures and snapshots presented in this paper use
\textsc{matplotlib}\cite{Matplotlib}.

\section{Validation with locally generated data}
\label{sec:LocalData}

In this section the memory-usage model parametrisation
and seasonal adjustment processes are tested and validated,
using the memory-usage data from a single application
and the network-traffic data from scheduled server requests, respectively.

\textbf{Memory usage parametrisation:} A total of 4 days of data is analysed.
The running time of the algorithm is approximately 1.5 minutes.

The model parameters extracted are well compatible with the features of the input data,
therefore validating the procedure. For $max\ memory$ and $\beta$ the distributions
have a negligible standard deviation.
The $run\ time$ parameter distribution is broader.
This could be due either to a limitation of the algorithm or to the derivatives-based strategy,
or that the run time of an application can not be described in terms of a single parameter.

\begin{figure}[!hbt]
  \begin{center}
    \includegraphics[width=\columnwidth]{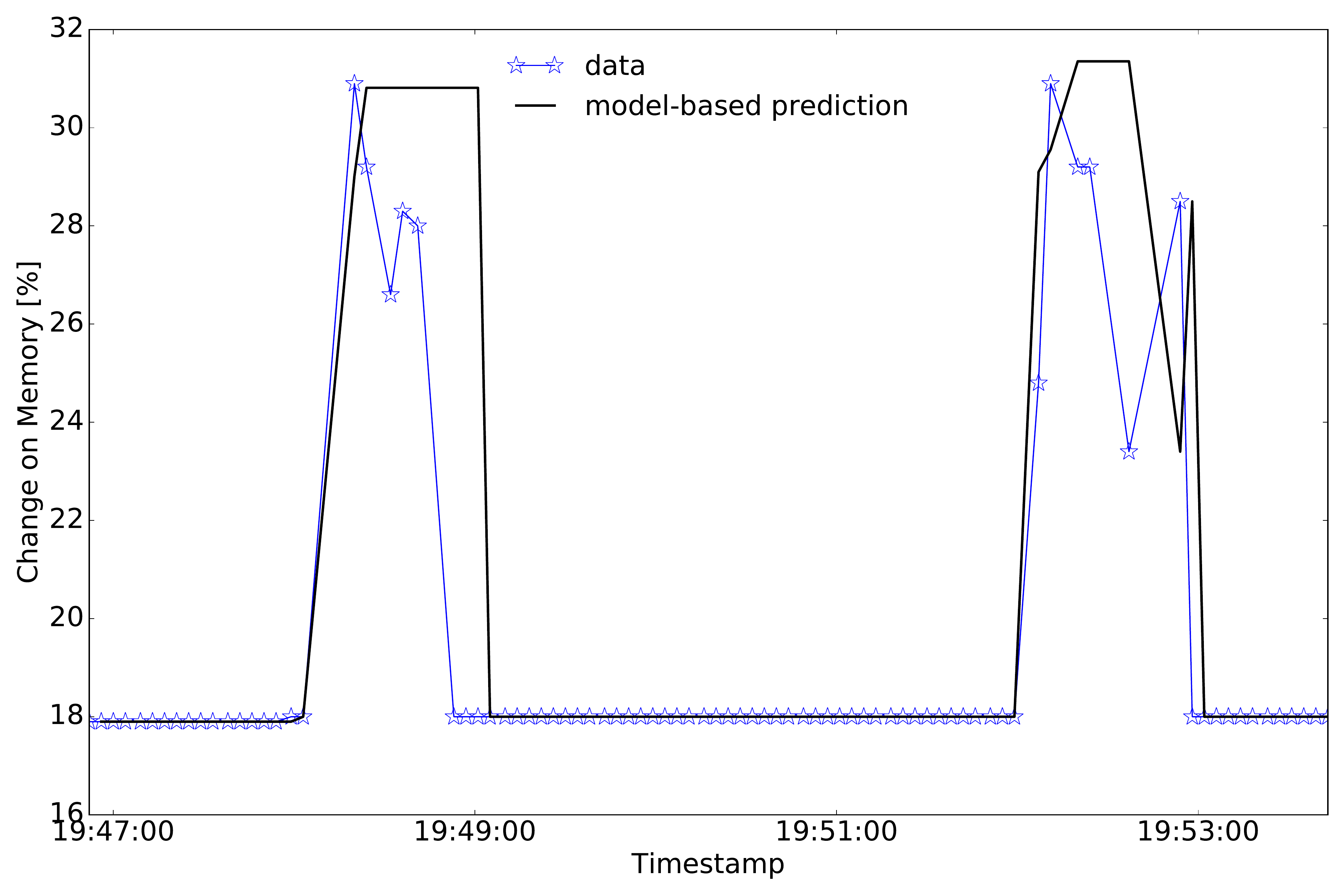}
    \caption{Data to model-prediction comparison for a small
      time window covering two signal-like patterns.}
    \label{fig:TriggerAndForecast}
  \end{center}
\end{figure}

Fig.~\ref{fig:TriggerAndForecast} shows a snapshot of the memory-usage
parametrisation model in action.
The model prediction for $y_t$ is $y_{t-1}$ unless a trigger is activated,
in which case the signal model takes over.
For this validation, the trigger is given by a jump in the CPU usage.
The signal model takeover lasts for a time lapse given by the $run\ time$ parameter.
The parameter $\beta$ describes the turn-on features well.
The maximum memory usage is also reasonably well described.
The running time and turn-off are less well described, however, overall,
the model does a good job in encapsulating the data as intended.

\textbf{Seasonal adjustment:} A total of 2 days of data is analysed.
The execution time of the seasonal adjustment process is approximately 20 seconds.
This includes the reading and formatting of the apache log files.

\begin{figure}[!hbt]
  \begin{center}
    \includegraphics[width=\columnwidth]{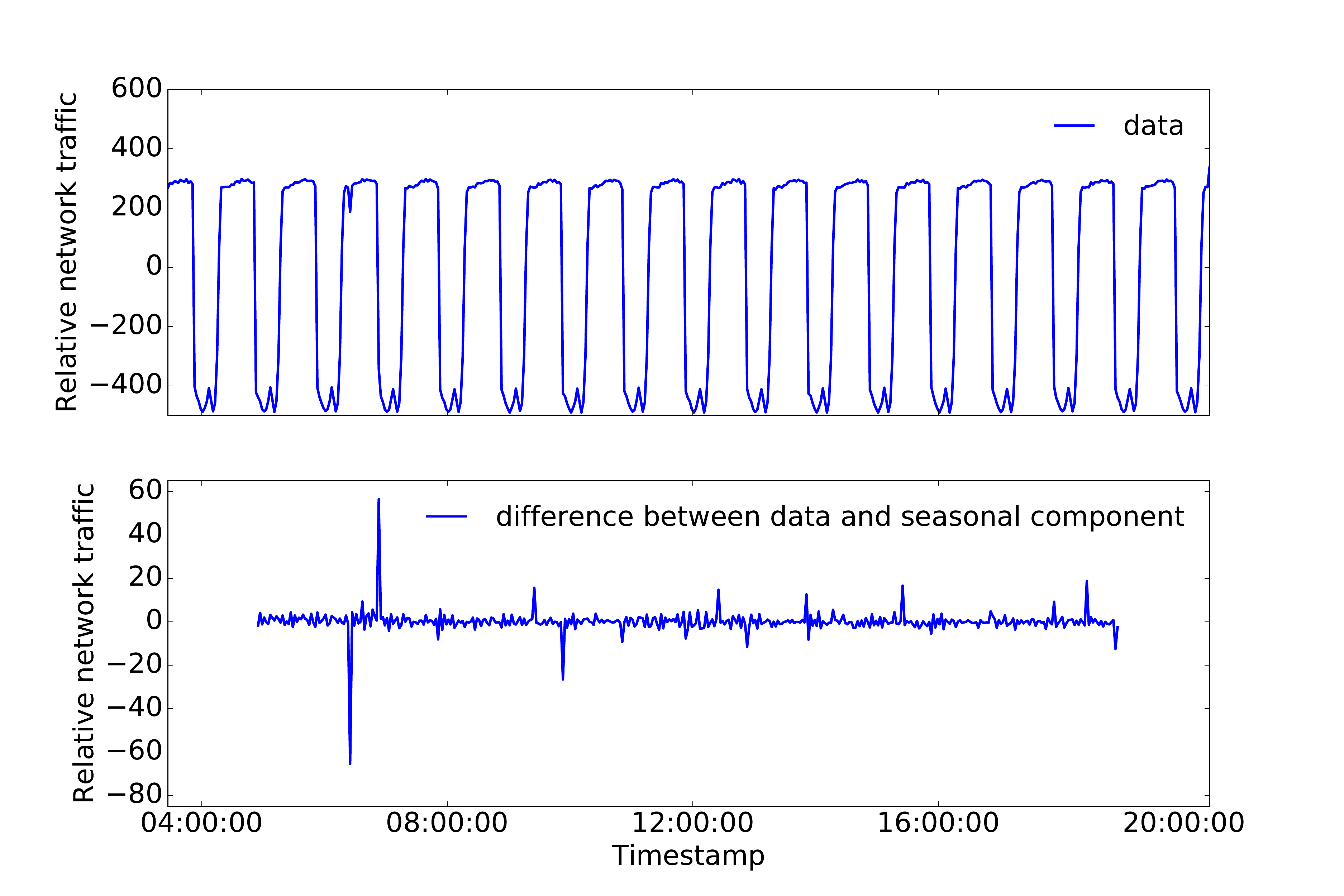}
    \caption{Top: Input data with a hourly seasonality.
      Bottom: residual difference between the input data and its seasonal component.
    }
    \label{fig:SeasonalDecomposition}
  \end{center}
\end{figure}

Fig.~\ref{fig:SeasonalDecomposition} shows the seasonal adjustment process in action.
The input data has an hourly periodicity by design.
The data is translated vertically using the MA,
so that it fluctuates around zero.
This step is not necessary for the seasonal adjustment to work, however,
it helps to illustrate the size of the residue, which is shown in the bottom panel.
This residue shows the difference between the input data and its seasonal component,
which, as expected, is small, demonstrating that the seasonal adjustment process works well.
The short-lived spikes observed are understood in terms of faulty lines in the apache log files,
causing holes in the data.
The missing data at the beginning and at the end of the residue are a consequence
of the usage of a MA for the seasonal adjustment process,
as discussed in Section~\ref{sec:Theory}.

\section{Using AUGURY for the analysis of web-server usage data}
\label{sec:RealData}

\begin{figure}[!hbt]
  \begin{center}
    \includegraphics[width=\columnwidth]{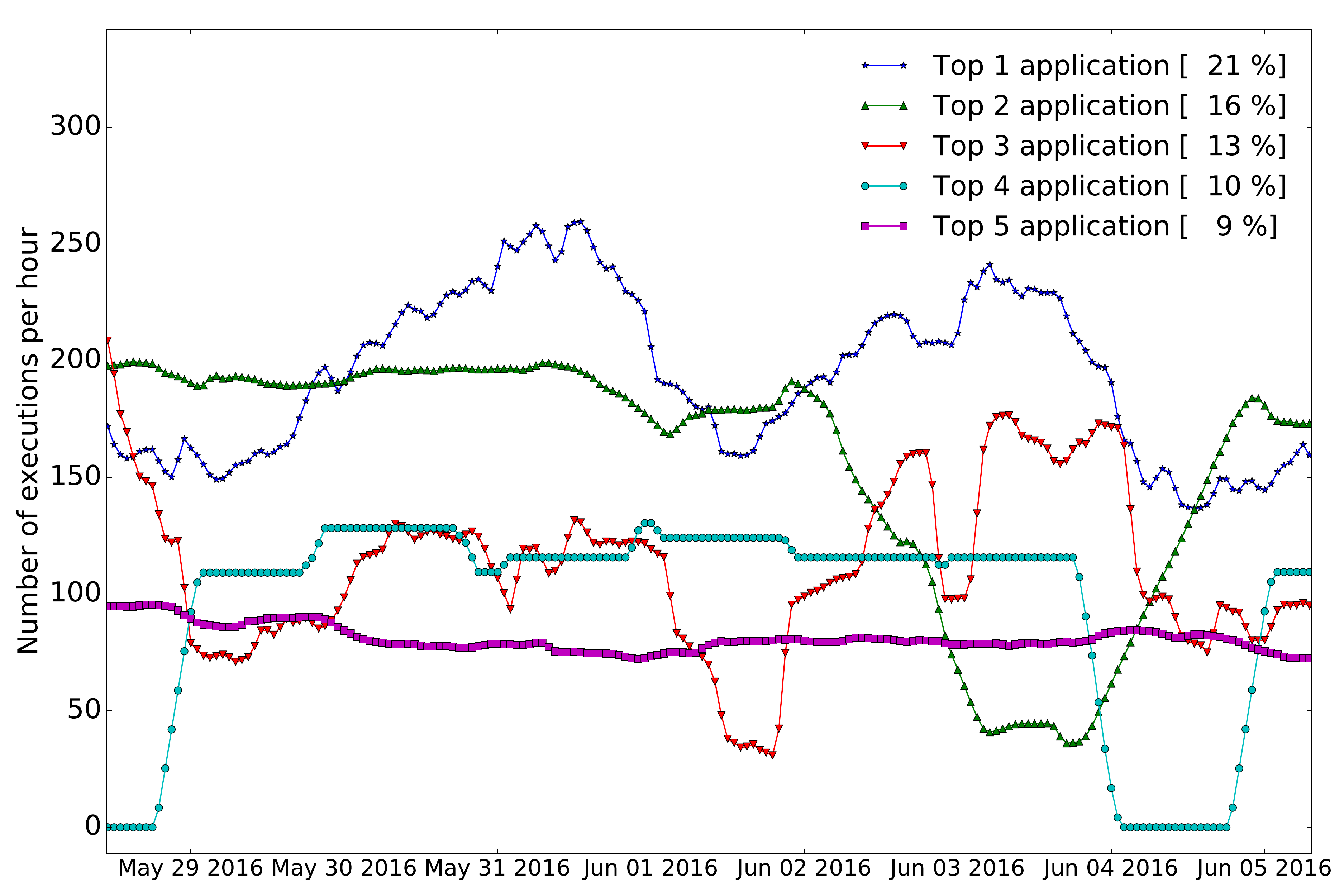}
    \caption{Trend component of each of the top 5 applications,
      ranked according to their share of the total traffic data.
      For each application, the seasonal, trend and residual components are extracted
      from the input data using seasonal adjustment.
      The data shown here corresponds to a period of 7 days, from Sunday to Sunday.}
    \label{fig:RealDataTrends}
  \end{center}
\end{figure}

The seasonal patterns are studied by looking at the number of executions per
application within a given time window.
AUGURY allows the user to set this window to any number of minutes.
Fig.~\ref{fig:RealDataTrends} shows the trend component of the input data
for the top 5 most frequently used applications.
Here, a time window of one hour is used.
The trend component represents the evolution of the mean network traffic.
Each application has a unique trend.
In some cases a flat behaviour is observed. In others,
hints of a weekly seasonality are displayed.
In the remaining cases, an accurate forecast for the trend component can hardly be achieved,
unless the force driving the trend is identified.
However, most of the sudden changes on the network traffic are absorbed
by the residual (stochastic error) term and, hence, that is the component which carries
the most potential damage to the server.

\begin{figure*}[tbp]
  \begin{center}
    \includegraphics[width=\textwidth]{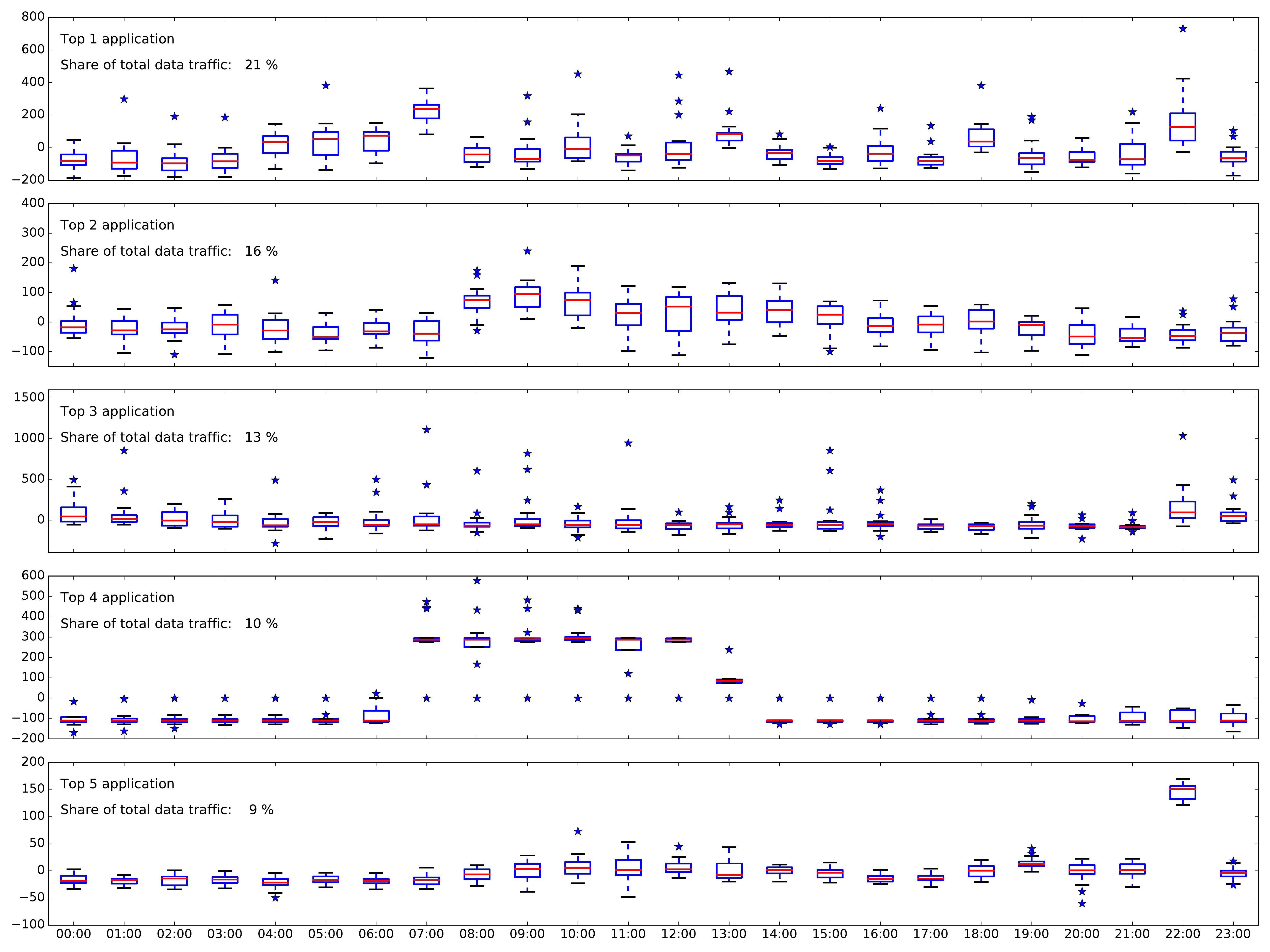}
    \caption{Daily seasonality of the top 5 applications,
      ranked according to their share of the total traffic data.
      The trend component has been subtracted using seasonal adjustment.
      For each hour there are as many entries
      as there are days in the input data.
      Each entry shows the total number of executions,
      relative to the trend component,
      for a given day at the
      hour indicated by the x-axis.
      In each sample, the red line shows the median (second quartile), 
      the bottom and top of the box are the first and third quartiles,
      and the ends of the whiskers show the lowest (highest) entry still
      within 1.5 interquartile ranges of the lower (upper) quartile.
      Outliers are represented as star-shaped markers. 
    }
    \label{fig:RealDataSeasonal}
  \end{center}
\end{figure*}

\begin{figure*}[tbp]
  \begin{center}
    \includegraphics[width=\textwidth]{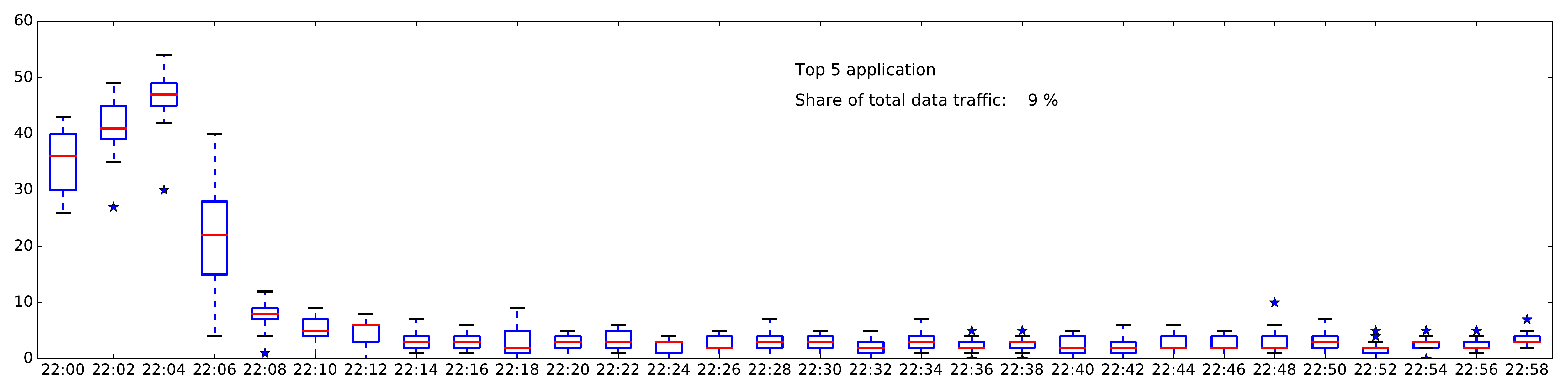}
    \caption{Hourly seasonality for the 5$^{\mathrm{th}}$ most frequently used application.
      This corresponds to a zoom of the daily seasonality plot,
      obtained by selecting entries occurring between $22:00$ and $23:00$.
      For each minute there are as many entries as there are days in the input data.
      Each entry shows the total number of executions, for a given day at chosen hour,
      as indicated in the x-axis.
      In each sample, the red line shows the median (second quartile), 
      the bottom and top of the box are the first and third quartiles,
      and the ends of the whiskers show the lowest (highest) entry still
      within 1.5 interquartile ranges of the lower (upper) quartile.
      Outliers are represented as star-shaped markers. 
    }
    \label{fig:RealDataSeasonalZoom}
  \end{center}
\end{figure*}

The size of the daily seasonal component
relative to the trend component varies from application to application.
Fig.~\ref{fig:RealDataSeasonal} shows this contribution
for the five most frequently used applications.
For the first application in the ranking, the median fluctuates between $+200$ and $-100$,
approximately,
around the average value of the trend component at $200$.
For the last application in the ranking,
the median of the seasonal component peaks at around $+150$,
which is a sizeable contribution for an application
with an average trend component of approximately $100$.
In Fig.~\ref{fig:RealDataSeasonal}, the seasonal component is also shown relative to the residual.
This is illustrated through the box and whisker,
encapsulating 50$\%$ of the data and $+1.5$ ($-1.5$)
interquartile ranges from the top (bottom) edges of the box, respectively.
A significant change on the seasonal pattern,
relative to the neighbouring points along the x-axis, is that which
is not covered by the overlapping boxes or whiskers.

A graphic such as Fig.~\ref{fig:RealDataSeasonal} presents abundant information
in a concise manner,
relevant to both diagnosis and forecasting purposes.
Issues can be tracked down to the source by searching for sudden increases
on the traffic data in a very narrow time window.
A systematic congestion is illustrated, for instance,
in the bottom panel of Fig.~\ref{fig:RealDataSeasonal}.
Extraordinary events can also be shown with respect to a forecast based on the seasonal pattern.
Outliers such as those indicated by the star-shaped markers in Fig.~\ref{fig:RealDataSeasonal}
can be easily identified, via this forecasting strategy.

A timely identification of an outlier or a precise pin-pointing of a systematic congestion
requires a more finely grained version of Fig.~\ref{fig:RealDataSeasonal}.
Functionality for this is provided by AUGURY.
Fig.~\ref{fig:RealDataSeasonalZoom} shows the hourly seasonality
for the peak usage hour of the application
shown in the bottom panel of Fig.~\ref{fig:RealDataSeasonal}.
Here, the traffic data is shown to be concentrated in the first
minutes of the $22:00-23:00$ period.

\begin{figure}[!hbt]
  \begin{center}
    \includegraphics[width=\columnwidth]{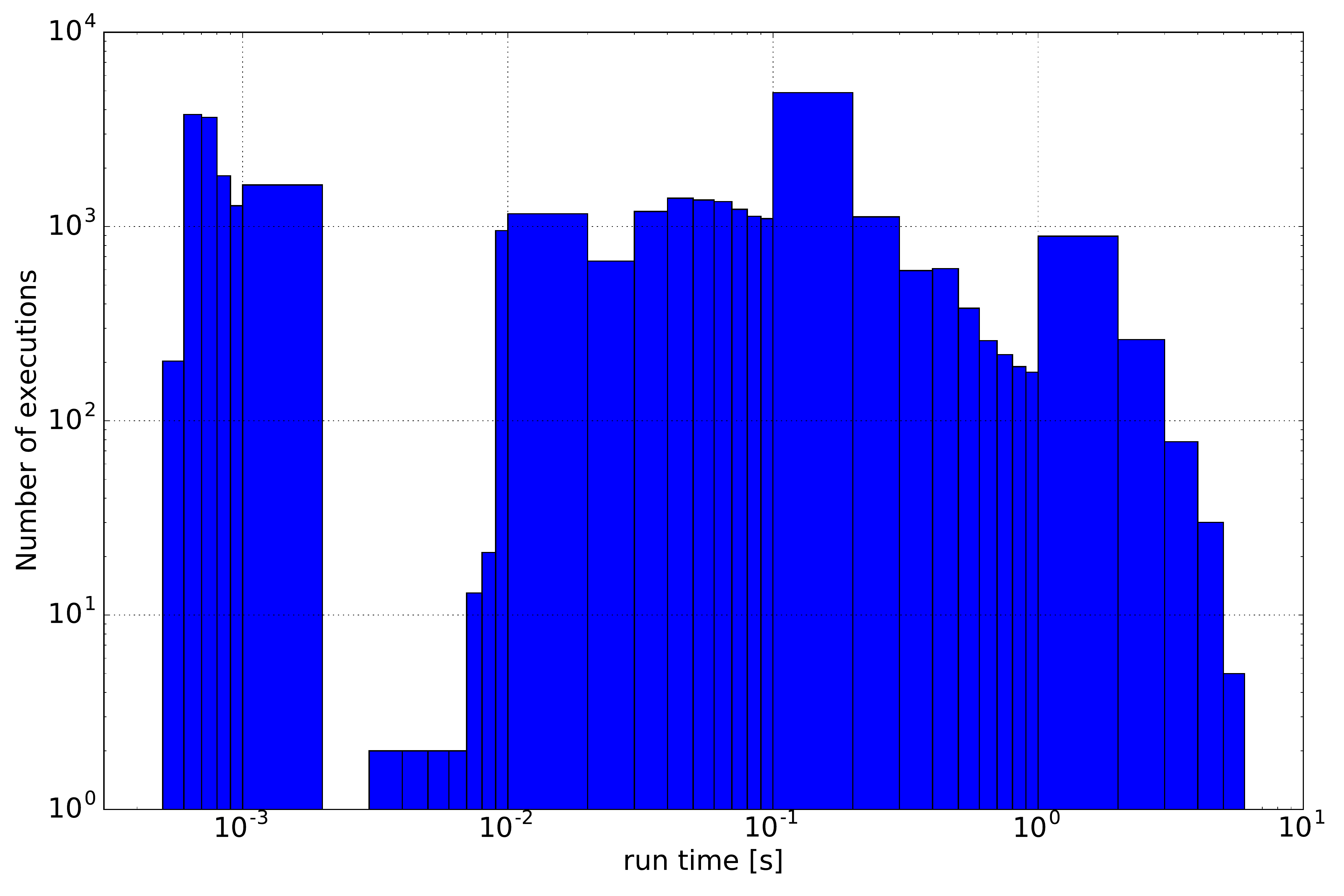}
    \caption{Run time for the 5$^{\mathrm{th}}$ most frequently used application.}
    \label{fig:RealDataRunTime}
  \end{center}
\end{figure}

The two essential ingredients for the model-based approach introduced in this paper
are single-valued maximum memory usage and run time per application.
The former realises in data, however, the latter is observed to behave differently.
Fig.~\ref{fig:RealDataRunTime} shows the run time for the application
shown in the bottom panel of Fig.~\ref{fig:RealDataSeasonal}.
Here, the run time is shown to fluctuate between milliseconds and seconds.
This fluctuation can be understood in terms of data availability.
A snapshot such as Fig.~\ref{fig:RealDataRunTime}
can be used by a system administrator to identify and diagnose bottlenecks.

\begin{figure}[!hbt]
  \begin{center}
    \includegraphics[width=\columnwidth]{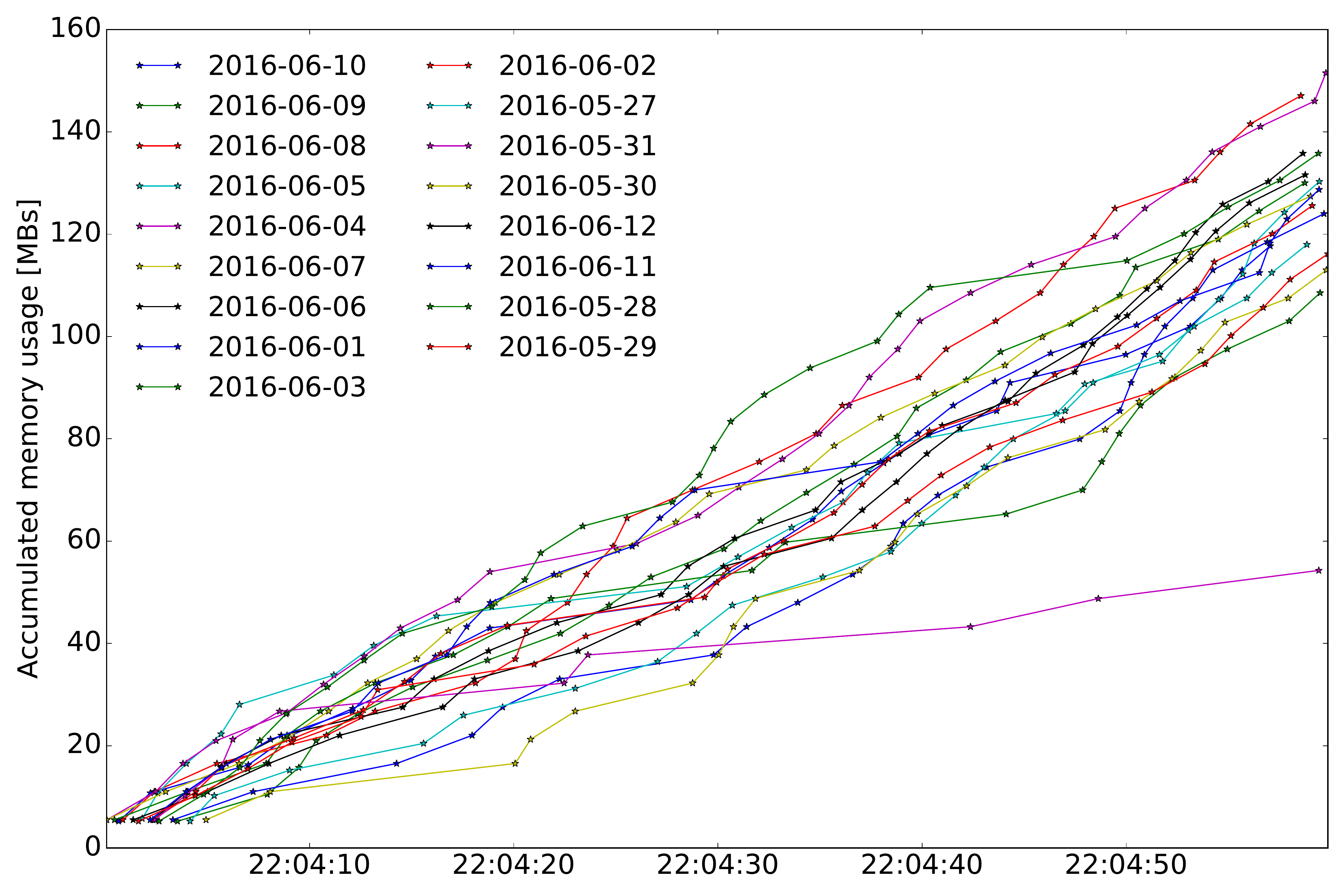}
    \caption{Accumulated memory usage for the 5$^{\mathrm{th}}$ most frequently used application,
      for a time window of one minute.
      This is shown separately for each of the days in the dataset.
      The x-axis shows the execution time.
      Here, a subset of the data is used, corresponding to a time window of one minute,
      between $22:04$ and $22:05$.
      This corresponds to the peak usage time.}
    \label{fig:ReadDataAcummMem}
  \end{center}
\end{figure}

The parametrisation of the memory usage via the model discussed in
Section~\ref{sec:Modelling} can be applied here,
despite the fact that the run time does not take a single value per application.
Fig.~\ref{fig:ReadDataAcummMem} shows the cumulative memory usage,
for a minute-sized time window,
corresponding to that of the peak usage time of the application.
This shows the worst-case scenario, where memory is not being released,
i.e. the run-time of the application extends beyond the time window.
For this scenario, an accumulated memory of 150 MBs is reached.
A graphic such as Fig.~\ref{fig:ReadDataAcummMem} can be used by
a system administrator to prevent system overloads
by revealing dangerous tendencies on the memory usage at peak operating times.

\section{Modelling of the residual component of web-server data}
\label{sec:Residual}

In this section, the auxiliary functionality that AUGURY provides
for the forecasting of the stochastic
fluctuations around the stable long-term prediction, given by the seasonal component,
is discussed.

The forecasting is based on the Auto Regressive Integrated
Moving Average~\cite{ARIMA1,ARIMA2} (ARIMA) model.
The parameters of the ARIMA model are:
\begin{itemize}
\item $p$, number of auto regressive terms;
\item $d$, number of differences, e.g. if $d=1$ is used a fit to $Y'$ is made;
\item $q$, number of moving average terms.
\end{itemize}
This model can only be applied to stationary series.
Stationarity is a property of those time series which are dominated
by random fluctuations around an stable long-term average.
This property implies that deviations from it are likely to be followed by a regression to it.

The stationarity of the residual term of the web-server data is tested using the
Augmented Dickey-Fuller\footnote{\url{http://faculty.smu.edu/tfomby/eco6375/BJ\%20Notes/ADF\%20Notes.pdf}} (ADF) test~\cite{ADF}.
The principle behind the ADF test can be illustrated in terms of
the correlation between two consecutive observations, $y_t$ and $y_{t-1}$.
In a simple AR(1) model, the difference between them is modelled as
\begin{equation}
  y'_t = \delta y_{t-1} + e_t.
\label{eq:AR}
\end{equation}
In Eq.~\ref{eq:AR}, $-2 < \delta < 0$ and $e_t$ is a zero-mean error term.
If $\delta = 0$, the process behaves as a random walk and, hence,
the difference does not converge nor does the average $y_t$ tend towards zero.
If $\delta = -1$, the mean of the difference is zero and, in general,
if $\delta \leq-1$, a zero-mean or change-of-sign-like behaviour is implied.
This latter case corresponds to that of stationary series.
The ADF test checks the $\delta = 0$ hypothesis.
The stronger the rejection of this hypothesis, the more likely the series is to be stationary.
The actual ADF model is more general, as it allows for a constant,
a trend and a correlation between differences at a lagged time.
This model is given by
\begin{equation}
  y'_t = \alpha + \beta t + \gamma y_{t-1} + \delta_1 y'_{t-1} + \dots + \delta_{p-1} y'_{t-p+1} + e_t
\end{equation}
Here, the $\gamma = 0$ hypothesis is tested.

The residual term of the web-server data is divided into two sets.
The first part is used as in-sample data for fitting the ARIMA model.
The second part is used as off-sample data, for which a forecast is made.
The forecast is performed in an iterative fashion.
First, a forecast is obtained for the first point of the in-sample data.
The actual value the series takes at that point is then added to the in-sample data
and subsequently removed from the off-sample data.
This continues until all the points on the off-sample data are exhausted
and a forecast for each exists.
The ARIMA model is fitted on each iteration.

\begin{figure*}[tbp]
  \begin{center}
    \includegraphics[width=\textwidth]{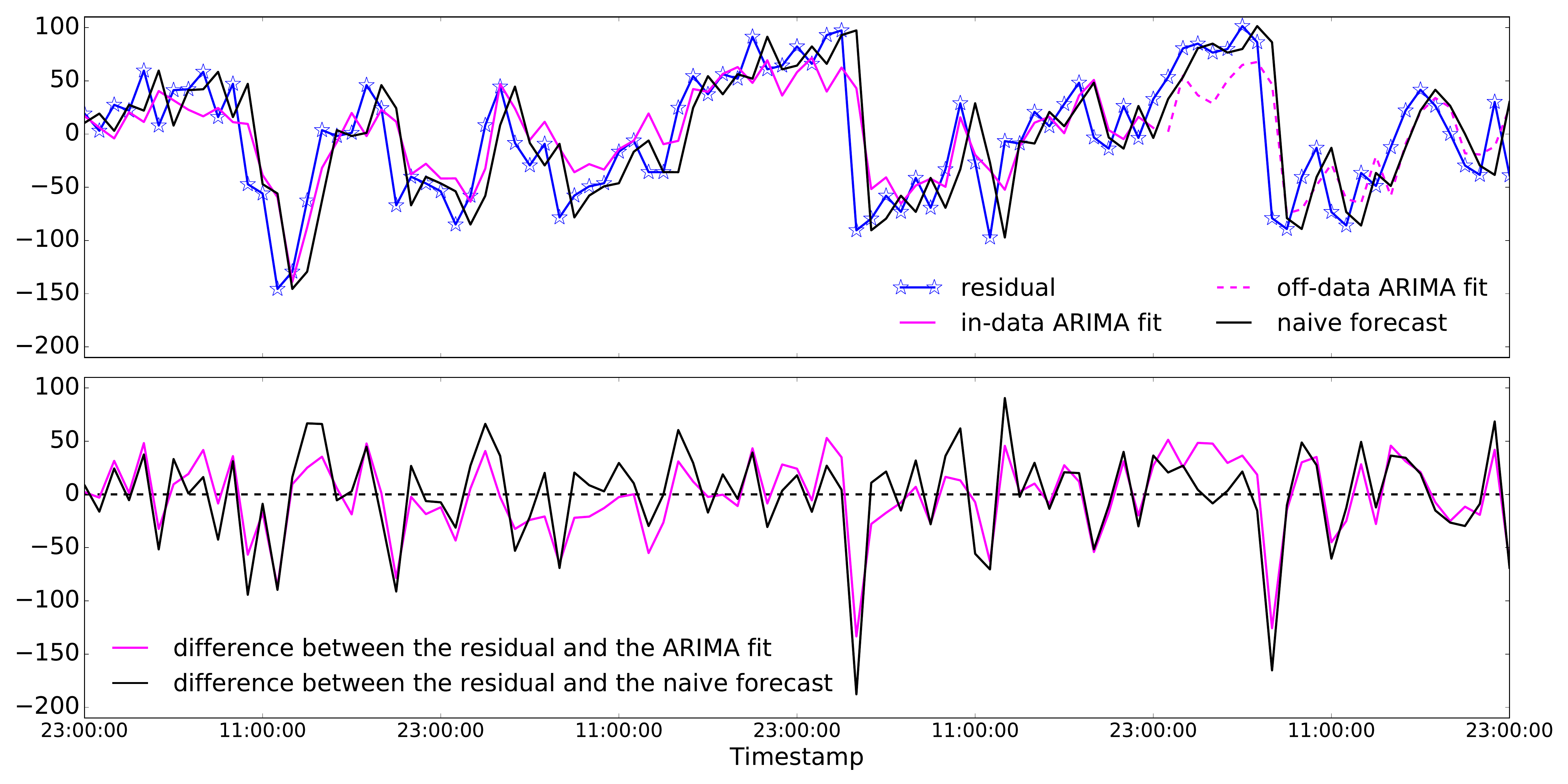}
    \caption{ARIMA fit of the residual term of the network
      traffic data after seasonal adjustment.}
    \label{fig:ARIMAFit}
  \end{center}
\end{figure*}

Fig.~\ref{fig:ARIMAFit} shows the results of the iterative ARIMA fit,
compared to a naive ($y_{t+1} = y_{t}$) forecast.
The ARIMA fit is shown not to surpass the benchmark set by the naive forecast.
In fact, they are rather compatible, meaning that, according to the ARIMA fit,
the residual behaviour is that of a random walk, $y_{t+1} = y_t + \epsilon_{t+1}$
where $\epsilon_{t+1}$ is a random increment with a zero mean.
This result does not come as a surprise, as previous work
has reached the same conclusion~\cite{Related6},
using sophisticated forecasting methods,
while studying data with a similar noise-like behaviour.

\section{Summary}
\label{sec:Summary}

AUGURY, an application for the analysis of monitoring data from computers, server
or cloud infrastructures has been presented.
Its components, a tool for memory-usage modelling and pattern extraction,
and a tool for the study of the seasonality in network traffic, have been validated.
The insight that AUGURY adds to the analysis of monitoring data
has been demonstrated via snapshots that make seasonal patterns apparent.

The use case for AUGURY is the study of seasonal patterns.
It has been shown how an administrator can, with a glance at AUGURY's output,
gauge the strength of the seasonal pattern with respect to the trend and residue components,
and easily identify a significant network traffic congestion
and peak usage times.
AUGURY provides functionality to study and diagnose these events further,
which was illustrated through memory-usage projections and run-time diagnosis.

AUGURY's approach to the aleatory component in data driven by human consumption
is also built around seasonal patterns.
Statistical outliers are made apparent,
leaving to the discretion of the user to judge whether individual attention is needed.
The remaining random fluctuations are embedded within the seasonal component,
instead of pursuing an accurate prediction.
In this work, it has been shown that using standard methods on this pursuit
is a fruitless effort. Thus, in AUGURY, this stochastic component is
instead used to indicate how likely the seasonal component is to take its median value,
to help measuring its strength relative to the trend and/or stochastic components.

\section*{Acknowledgements}
The authors would like to thank Bashar Ahmad, Paul Heinzlreiter and Michael Krieger
for their suggestions and valuable discussion.
The work of N.G. is supported by the 7th Framework Programme of the European
Commission through the Initial Training Network HiggsTools
PITN-GA-2012-316704.

\end{document}